\documentstyle[psfig]{mn}

\newif\ifAMStwofonts

\def\mst{\begin{displaymath}}
\def\mend{\end{displaymath}}
\def\tcool{t_{\rm cool}}
\def\rcool{r_{\rm cool}}
\def\tflow{t_{\rm flow}}

\def\mh{m_{\rm H}}
\def\me{m_{\rm e}}
\def\vrot{v_{\rm rot}}

\def\ne{n_{\rm e}}
\def\nh{n_{\rm H}}
\def\vb{v_{\rm B}}
\def\rx{r_{\rm x}}

\def\muh{\mu_{\rm h}}
\def\tg{t_{\rm BH}}
\def\tgc{t_{\rm BH,c}}
\def\thf{t_{\rm am}}
\def\tf{t_{\rm d}}
\def\mdedd{\dot M_{\rm Edd}}
\def\alphah{\alpha_{\rm h}}
\def\alphad{\alpha_{\rm d}}
\def\rb{r_{\rm B}}
\def\etal{{et al\thinspace}}

\def\eg{{\em e.g.\ }}

\def\ie{{\em i.e.\ }}
\def\spose#1{\hbox to 0pt{#1\hss}}
\def\approxlt{\mathrel{\spose{\lower 3pt\hbox{$\sim$}}
        \raise 2.0pt\hbox{$<$}}}
\def\approxgt{\mathrel{\spose{\lower 3pt\hbox{$\sim$}}
        \raise 2.0pt\hbox{$>$}}}   
\def\Mdot{\hbox{$\dot M$}}
\def\today{\ifcase\month\or January\or February\or March\or April\or May\or
      June\or July\or August\or September\or October\or November\or December\fi
      \space\number\day, \number\year}

\def\boxit#1{\vbox{\hrule\hbox{\vrule\kern3pt\vbox{\kern3pt
          #1 \kern3pt}\kern3pt\vrule}\hrule}}

\def\Msun{\hbox{$\rm\thinspace M_{\odot}$}}

\def\tcoll{t_{\rm coll}}
\def\eqst#1{\begin{equation} \label{#1}}
\def\eqend{\end{equation}}
\def\nh{n_{\rm H}}

\def\half{{1\over2}}
\def\machi{{\cal M}_{\rm i}}
\def\machin{{\cal M}_{\rm i,0}}
\def\ri{r_{\rm i}}
\def\Ti{T_{\rm i}}
\def\Mdoti{\dot M_{\rm i}}
\def\Sigi{\Sigma_{\rm i}}
\def\Mdotbh{\dot M_{\rm h}}
\def\Mbh{M_{\rm h}}

\def\rhoi{\rho_{\rm i}}
\def\si{s_{\rm i}}
\def\vi{v_{\rm i}}
\def\rcf{r_{\rm CF}}
\def\sigmat{\sigma_{\rm T}}

\ifoldfss
  \ifCUPmtlplainloaded \else
    \NewTextAlphabet{textbfit} {cmbxti10} {}
    \NewTextAlphabet{textbfss} {cmssbx10} {}
    \NewMathAlphabet{mathbfit} {cmbxti10} {} 
    \NewMathAlphabet{mathbfss} {cmssbx10} {} 
  \fi
  \ifAMStwofonts
    \ifCUPmtlplainloaded \else
      \NewSymbolFont{upmath} {eurm10}
      \NewSymbolFont{AMSa} {msam10}
      \NewMathSymbol{\upi}     {0}{upmath}{19}
      \NewMathSymbol{\umu}     {0}{upmath}{16}
      \NewMathSymbol{\upartial}{0}{upmath}{40}
      \NewMathSymbol{\leqslant}{3}{AMSa}{36}
      \NewMathSymbol{\geqslant}{3}{AMSa}{3E}

    \fi
  \fi
\fi 

\ifnfssone
  \newmathalphabet{\mathit}
  \addtoversion{normal}{\mathit}{cmr}{m}{it}
  \addtoversion{bold}{\mathit}{cmr}{bx}{it}
  \newmathalphabet{\mathbfit} 
  \addtoversion{normal}{\mathbfit}{cmr}{bx}{it}
  \addtoversion{bold}{\mathbfit}{cmr}{bx}{it}
  \newmathalphabet{\mathbfss} 
  \addtoversion{normal}{\mathbfss}{cmss}{bx}{n}
  \addtoversion{bold}{\mathbfss}{cmss}{bx}{n}
  \ifAMStwofonts
    \ifCUPmtlplainloaded \else
      \UseAMStwoboldmath
      \makeatletter
      \new@mathgroup\upmath@group
      \define@mathgroup\mv@normal\upmath@group{eur}{m}{n}
      \define@mathgroup\mv@bold\upmath@group{eur}{b}{n}
      \edef\UPM{\hexnumber\upmath@group}
      \new@mathgroup\amsa@group
      \define@mathgroup\mv@normal\amsa@group{msa}{m}{n}
      \define@mathgroup\mv@bold\amsa@group{msa}{m}{n}
      \edef\AMSa{\hexnumber\amsa@group}
      \makeatother
      \mathchardef\upi="0\UPM19
      \mathchardef\umu="0\UPM16
      \mathchardef\upartial="0\UPM40
      \mathchardef\leqslant="3\AMSa36
      \mathchardef\geqslant="3\AMSa3E
    \fi
  \fi
\fi 

\ifnfsstwo
  \DeclareMathAlphabet{\mathbfit}{OT1}{cmr}{bx}{it}
  \SetMathAlphabet\mathbfit{bold}{OT1}{cmr}{bx}{it}
  \DeclareMathAlphabet{\mathbfss}{OT1}{cmss}{bx}{n}
  \SetMathAlphabet\mathbfss{bold}{OT1}{cmss}{bx}{n}
  \ifAMStwofonts
    \ifCUPmtlplainloaded \else
      \DeclareSymbolFont{UPM}{U}{eur}{m}{n}
      \SetSymbolFont{UPM}{bold}{U}{eur}{b}{n}
      \DeclareSymbolFont{AMSa}{U}{msa}{m}{n}
      \DeclareMathSymbol{\upi}{0}{UPM}{"19}
      \DeclareMathSymbol{\umu}{0}{UPM}{"16}
      \DeclareMathSymbol{\upartial}{0}{UPM}{"40}
      \DeclareMathSymbol{\leqslant}{3}{AMSa}{"36}
      \DeclareMathSymbol{\geqslant}{3}{AMSa}{"3E}
    \fi
  \fi
\fi 

\ifCUPmtlplainloaded \else
  \ifAMStwofonts \else 
    \def\upi{\pi}
    \def\umu{\mu}
    \def\upartial{\partial}
  \fi
\fi

\long\def\boxit#1{\noindent\ignorespaces
  \framebox[\hsize][l]{\hbox{\vbox{\raggedright #1\par}}}\par
  \medskip\noindent\ignorespaces
} 

\title{Fuelling quasars with hot gas}

\author[Nulsen \& Fabian]
{P.E.J.~Nulsen,$^{1,2}$ and A.C.~Fabian$^1$\\
$^1$Institute of Astronomy, Madingley Road, Cambridge CB3 0HA \\
$^2$Department of Physics, University of Wollongong,
Wollongong NSW 2522, Australia \\}

\date{9 July 1999}
\pagerange{\pageref{firstpage}--\pageref{lastpage}}      
\pubyear{1999}

\begin{document}

\maketitle

\label{firstpage}

\begin{abstract}
We consider a model for quasar formation in which massive black holes are
formed and fuelled largely by the accretion of hot gas during the process of
galaxy formation.  In standard hierarchical collapse models, objects about
the size of normal galaxies and larger form a dense hot atmosphere when they
collapse.  We show that if such an atmosphere forms a nearly ``maximal''
cooling flow, then a central black hole can accrete at close to its
Eddington limit.  This leads to exponential growth of a seed black hole,
resulting in a quasar in some cases.  In this model, the first quasars form
soon after the first collapses to produce hot gas.  The hot gas is depleted
as time progresses, mostly by cooling, so that the accretion rate eventually
falls below the threshold for advection-dominated accretion, at which stage
radiative efficiency plummets and any quasar turns off.  A simple
implementation of this model, incorporated into a semi-analytical model for
galaxy formation, over-produces quasars when compared with observed
luminosity functions, but is consistent with models of the X-ray Background
which indicate that most accretion is obscured. It produces few quasars at
high redshift due to the lack of time needed to grow massive black holes.
Quasar fuelling by hot gas provides a minimum level, sufficient to power
most quasars at redshifts between one and two, to which other sources of
fuel can be added. The results are sensitive to feedback effects, such as
might be due to radio jets and other outflows.
\end{abstract}


\section{Introduction} 

It is generally accepted that quasars are the result of accretion onto
massive black holes residing in the nuclei of normal galaxies. Rees (1984)
has argued that a black hole is likely to form at the centre of almost any
galaxy, so that the main issue for quasar formation is how the black hole is
fuelled.  Models for quasar evolution must account for the time dependence
of the quasar luminosity function, particularly its peak at $z \sim 1.5$ and
subsequent decline (\eg Boyle, Shanks \& Peterson, 1988), and also for the
formation of the massive black holes and their remnants that reside in the
nuclei of many nearby galaxies (\eg Ford \etal 1997; Magorrian \etal 1998).

A variety of models has been proposed for the fuelling of quasars,
most of which rely on making interstellar gas fall close to the
nucleus where it joins an accretion disc (\eg Shlosman, Begelman \&
Frank 1990).  In this paper we consider the possibility that the main
source of fuel is the hot interstellar medium formed during the
collapse of larger galaxies.  Provided that the angular momentum of
the gas is not too large, a nuclear black hole can grow by Bondi
accretion and, if the gas temperature is close to the virial
temperature, its growth time is controlled largely by the density of
the hot gas.  In section 3 we show that, soon after the collapse of a
protogalaxy, this can be large enough to make the accretion rate of a
nuclear black hole comparable to the Eddington rate.

It is the protogalaxies with the largest spheroids, \ie large
elliptical protogalaxies, that contain most hot gas (Nulsen \& Fabian
1997), making them the best hosts for quasars in this model.  Thus,
the observation that quasars reside in elliptical hosts (McLure \etal
1998) is consistent with them being fuelled by hot gas.  Based on the
presence of companions close to many quasars, it has been argued that
gravitational interaction plays a significant role in the quasar
phenomenon (\eg Bahcall \etal 1997).  However, we note that this may
also be interpreted as indicating recent collapse.  As outlined below,
the supply of hot gas is expected to be greatest immediately after a
collapse and to decrease with time, so this would also be consistent
with fuelling quasars by hot gas.  Of course, active galactic nuclei,
including quasars, may obtain their fuel from a variety of sources.
We consider that hot gas provides a minimum fuelling rate, which can
be supplemented by cold gas, if available.

Cooling depletes the hot gas throughout the galaxy, so that the nuclear
accretion rate decreases with time after a protogalaxy collapses. The
depletion of the hot gas does not however simply explain the lack of
luminous quasars at the current epoch. The central black holes in nearby
elliptical galaxies are still immersed in accretable hot gas, yet generally
have low accretion luminosities (Fabian \& Canizares 1988; Di Matteo \&
Fabian 1997). As an example, one of the best candidates for a massive
nuclear black hole is M87 (Harms \etal 1994; Marconi \etal 1997).  A
solution to this problem has been argued by Di Matteo \& Fabian (1997) and,
specifically for M87, by Reynolds \etal (1996) in which the accretion flow
becomes advection-dominated (i.e. an ADAF, Narayan \& Yi 1995), so having a
low accretion efficiency. We adopt that hypothesis here and assume that when
the nuclear accretion rate falls below the threshold for ADAF formation most
quasars fade rapidly (Fabian \& Rees 1995; Yi 1996).

In section 4 we describe the incorporation of a simple version of this
model for quasar formation into a semi-analytical model for galaxy
formation (Nulsen \& Fabian 1997; see also Haehnelt \& Rees 1993 for
models relating quasar evolution to galaxy formation).  This is used
to show that the model can account for the broad features of the
history of quasars in section 5.  Section 6 has a brief discussion of
feedback on the model, particular that due to Compton cooling.  In
section 7 we discuss the limitations of the semi-analytical model for
quasar formation and some predictions of the model.  Our conclusions
are summarized in section 8.

\section{Angular momentum and the feeding of a nuclear black hole}

Shlosman \etal (1990) discuss the issues of getting a large mass of gas to
accrete into the nucleus of a galaxy.  Almost all galaxies have appreciable
net rotation, so that the main difficulty is the dissipation of angular
momentum, essentially all of which must be dissipated in order for gas to
accrete into a nuclear black hole. The total accreted matter should also
account for (most of) the mass of the nuclear black hole, which exceeds
$10^9\ \Msun$ in many cases (Ford \etal 1997; Magorrian \etal 1998), so that
the gas needs to be drained from a large region around the nucleus.  The
larger this region, the greater the difficulty of dissipating angular
momentum.

According to the standard argument, the effective viscosity in an
accretion disc can be expressed as (Shakura \& Sunyaev 1973)
\eqst{alpha}
\mu_{\rm d} = {\alphad \rho s^2 \over \Omega},
\eqend
where $\rho$ is the density and $s = \sqrt{kT/(\mu \mh)}$ is the
isothermal sound speed of gas in the disc, and 
\mst
\Omega(r) = \sqrt{GM(r) \over r^3}
\mend
is the angular frequency of a circular orbit.  The dimensionless
parameter $\alphad$ cannot normally exceed 1 and is generally thought
to be $\sim 0.1$ -- 0.3 (e.g. Cannizzo 1993).  The same parametrization can be
applied to the hot gas, \ie gas at about the virial temperature, but
then it is more appropriate to express the viscosity in terms of the
scale height, $w$,
\eqst{hot}
\muh = \alphah \rho s w.
\eqend
The cold gas moves on almost circular orbits, so that the time
required for gas from radius $r$ to drain to the centre of the disc is
roughly (\eg Shlosman \etal 1990)
\eqst{drain}
\tf = {r \over v_r} = {r \vrot \over \alphad \eta^2 s^2},
\eqend
where $v_r$ is the radial speed, $\vrot = r \Omega$ is the rotation
speed of the gas disc and $\eta = d\ln \Omega/d\ln r$ ($-3/2$ for a
keplerian disc).

Since the hot gas is pressure supported, it can drain much faster
than a cold disc.  However, angular momentum may still prevent it
from accreting directly onto a nuclear black hole.  For gas at about
the virial temperature, the scale height is $w \simeq r$ and the
speed of sound is close to the kepler speed, $\vrot$, at the same
radius.  A rough estimate of the time required to dissipate the
angular momentum of the hot gas is then 
\mst
\thf \simeq {r \over \alphah \vrot}.  
\mend
For $\alphah$ of order unity, this is comparable to the dynamical
time, while 
\mst {\thf\over \tf} \simeq
{\alphad s^2 \over \alphah \vrot^2} \ll 1,
\mend
if the disc gas is cold.

Thus, while the drainage of a cold accretion disc is governed by the
dissipation of angular momentum, it is only when hot gas flows inward
at speeds approaching the free-fall velocity that we need to be
concerned with the effects of rotation on the flow.  Dissipation of
angular momentum is less of a problem for the accretion of hot gas
than it is for cold gas.  

For flow speeds comparable to the speed of sound or faster, the angular
momentum of the hot gas will be largely conserved, so that its residual
angular momentum will cause it to eventually join a disc.  We assume
that this occurs at a sufficiently small radius to make the drainage
time of the disc from the point where the gas joins it point short.

\section{Feeding by hot gas} \label{feed}

In the collapse of small protogalaxies, radiative cooling is faster
than shock heating, so that gas ends up cold immediately after the
collapse (Rees \& Ostriker 1977; White \& Frenk 1991).  In larger
systems, which are more tenuous and have higher virial temperatures,
some of the gas can form a hot atmosphere after the collapse.  The
condition for the gas at radius $r$ to be part of a hot atmosphere is
that its radiative cooling time be longer than the free-fall time from
$r$ to the centre of the protogalaxy.  This cooling time is still
significantly smaller than the time at which the system collapses, so
that the hot gas will start to cool, forming a cooling flow (Fabian
1994), almost immediately after collapse.  A central black hole can
accrete hot gas from the central region of the cooling flow.

Based on observations of clusters of galaxies, we expect gas taking
part in the cooling flow to be sufficiently inhomogeneous to lead to
widespread thermal instability (Nulsen 1986; 1988).  The general
solution for an inhomogeneous cooling flow is complex, since the flow
of each phase must be tracked separately.  However, Nulsen (1986) has
argued that gas blobs moving relative to the mean flow tend to be
rapidly disrupted until they are small enough to be pinned to the mean
flow.  As a result, the phases tend to flow inward at approximately
the same speed, \ie to comove.  In the central part of the cooling
flow, where conditions change slowly relative to the flow time,
we can also expect the flow to be nearly steady.

For the purpose of simulating quasar formation, we take the potentials
of galaxies to be exactly isothermal.  In that case, the mean
temperature of the inhomogeneous gas mixture in a steady, comoving
cooling flow will be close to the constant virial temperature, the
exact relationship depending on details of the inhomogeneous density
distribution in the gas.  There is a class of self-similar,
inhomogeneous cooling flow solutions in which the mean gas
temperature is a constant multiple of the virial temperature, the
isothermal cooling flows (Nulsen 1998).  The mean gas density and
temperature in an isothermal cooling flow are related to the flow
time, $r/v$, by
\eqst{igc} 
{r\over v} = K {kT \over \mu \mh}{ \rho\over \ne\nh \Lambda(T)},
\eqend
where $r$ is the radius, $v$ the flow velocity, $\rho$ the mean gas density,
$\ne$ the mean electron number density and $\nh$ the mean hydrogen number
density.  $T$ is the effective gas temperature (defined so that the pressure
is $\rho kT/(\mu\mh)$) and $\mu\mh$ is the mean mass per gas particle.  The
dimensionless constant, $K$, depends on flow details, including the cooling
function and the radial dependence of the mass deposition rate.  In clusters
the mass flow rate, $\Mdot$, is found to be approximately proportional to
$r$ (Fabian 1994; Peres et al 1998), which we take to be exact for the
purpose of our model.  In that case, if the cooling function is approximated
by a power law, $\Lambda(T) \propto T^a$, then $K$ depends weakly on the
exponent $a$, ranging from 2.32 to 2.92 for $a$ in the range $-0.5$ to 0.5.
We adopt the representative value $K = 2.5$ for our calculations.

In an isothermal cooling flow with mass flow rate $\Mdot \propto r$,
the flow velocity is constant, regardless of the details of the
cooling function, so that the Mach number is also constant.  The
arguments below rely on the Mach number of the cooling flow being
close to unity initially, since this maximizes the density of the hot
gas in the vicinity of a nuclear black hole.  This is the most
critical aspect of the cooling flow model for quasar formation, since
it determines the Bondi accretion rate of the nuclear black hole.

When the Mach number of a cooling flow is low, the linear growth of
thermal instability is weak (Balbus \& Soker 1989).  The
inhomogeneous, isothermal cooling flow model relies on non-linear
thermal instability (contrary to a common misconception, very small
amplitude density fluctuations become non-linear in a cooling flow;
Nulsen 1997).  However, for Mach numbers of order unity, linear
thermal instability is much stronger.  If, contrary to our
assumptions, the gas is very nearly homogeneous and the thermal
instability is weak (or if $\Mdot \propto r^\eta$ with $\eta < 1$),
the Mach number of the cooling flow increases inward and, as it
approaches unity, thermal instability becomes strong, causing
widespread deposition of cold gas.  This tends to make the Mach number
saturate close to one, so that our assumption that the Mach number of
the cooling flow in the vicinity of the nucleus is close to one is not
sensitive to our assumptions.  A cooling flow with a Mach number
of order unity is simply a maximal cooling flow.  Conveniently, the
nuclear accretion rate can be determined exactly for the isothermal
cooling flow model (see below), but our quasar formation model is not
critically dependent on the assumption that the hot gas forms an
isothermal cooling flow.  Also note that the gas density in an
isothermal cooling flow with Mach number close to one is similar to
that obtained by other arguments for the maximum gas density in a
protogalaxy (Fall \& Rees 1985).

We assume that a nuclear black hole accretes any hot gas coming within
its influence by Bondi accretion.  At very small $r$ the residual
angular momentum causes the accreting gas to pass through a shock
(assumed to be radiative) and join an accretion disc.  Thus, the final
stages of the accretion are still assumed to be through a disc, but we
take the nuclear accretion rate to equal the Bondi rate.  Thus, the
nuclear accretion rate is determined by the density and temperature of
the hot gas at the point that the influence of the black hole becomes
dominant.  There are a number of reasons why this assumption may not
be valid, but we have adopted it as the simplest possibility.

The result (\ref{igc}) shows that the steady cooling flow is governed by
the requirement that the cooling time equals the flow time, to within
factors of order unity.  As cooling gas comes under the influence of
the black hole, its flow velocity will increase, reducing $r/v$ to the
point that cooling is no longer effective.  Thus, the transition
between the cooling flow and the Bondi solution occurs at about the
radius where the initial Mach numbers of the two flows are equal.

The Bondi accretion rate for a monatomic gas (with $\gamma = 5/3$) is
(Shu 1991)
\eqst{bondi}
\Mdotbh = \pi\rhoi {G^2 \Mbh^2 \over \si^3},
\eqend
where $\rhoi$ is the gas density and $\si$ the adiabatic sound
speed at large $r$, and $\Mbh$ is the mass of the black hole.  Well
outside the accretion radius, the density in the Bondi solution is
almost constant so that the velocity, $\vb$, can be determined from
the accretion rate, $\Mdotbh = 4\pi\rhoi \vb r^2$.  The gas
temperature is also nearly constant, so that Mach number is given
approximately by
\mst
{\vb\over \si} \simeq {\Mdotbh \over 4\pi \rhoi \si r^2} 
= \left(G\Mbh \over 2\si^2 r\right)^2.
\mend
Equating this to the Mach number, $\machi$, of the cooling
flow, we find that the Bondi solution takes over at about the radius
\eqst{rx}
\rx = {G\Mbh\over 2 \si^2 \machi^{1/2}}.
\eqend

Using (\ref{igc}) for the gas density and (\ref{rx}) to replace $r$ in
the result, we can evaluate the accretion rate (\ref{bondi}) as
\eqst{mdot}
\Mdotbh = 2 \pi K \machi^{3/2} {k\Ti G\Mbh \over \mu\mh \Lambda(\Ti)}
{\rho^2\over \ne\nh},
\eqend
where $\Ti$ is the gas temperature at large $r$ (note that the
last factor is a constant).  

This result is shown to be exact in the Appendix, where $\Ti$ is now
to be interpreted as the ``mean'' gas temperature of a steady isothermal
cooling flow and $\machi$ as its Mach number.  The argument in the
Appendix also shows that, for the conditions of our model, \ie gas
conditions that would give an isothermal cooling flow with $\Mdot
\propto r$, the nuclear accretion rate is independent of the
gravitational potential in between the edge of the steady cooling flow and
the nucleus.  Thus, the accretion rate is largely unaffected by changes to
the potential of the galaxy due to deposition of cooled gas. This feature of
the model reflects an accidental balance between the competing effects of
deepening the potential: a temperature rise, reducing the Bondi accretion
rate, and a density increase, tending to increase it.

The result (\ref{mdot}) shows that a black hole at the centre of a
protogalaxy grows exponentially by Bondi accretion.  The timescale for
growth is
\eqst{growth}
\tg = {\Mbh\over\Mdotbh} = {1\over 2 \pi K} {\ne\nh\over\rho^2}
{\mu\mh\Lambda(\Ti) \over k \Ti G \machi^{3/2}},
\eqend
which depends only on the gas temperature and the Mach number in the
outer parts of the of the steady cooling flow.  Numerically, 
\eqst{ngrow}
\tg \simeq 5.1\times10^8 \Lambda_{-23} T_6^{-1} \machi^{-3/2}
{\rm\, y},
\eqend
where $\Ti = 10^6 T_6$ K and $\Lambda(\Ti) = 10^{-23} \Lambda_{-23}
{\rm\, erg\, cm^3\,  s^{-1}}$.

This can be compared directly with the growth timescale of a black hole
accreting at the Eddington rate, $t_{\rm Edd}$, as
\mst
\tg \simeq 11 \Lambda_{-23} T_6^{-1} \machi^{-3/2} t_{\rm Edd}.
\mend
Thus, for the relevant temperatures, the accretion rate from a maximal
cooling flow is about one tenth or more of the Eddington rate.

The Bondi radius for a stellar mass object is very small,
\mst 
\rb = {G\Mbh \over \si^2} \simeq 6\times10^{11} \Mbh T_6^{-1}
{\rm\ cm},
\mend 
for $\Mbh$ in solar masses, so we should be wary of applying our result
to accretion onto very small black holes.  However, we assume that
more massive seed black holes are formed in the cores of all galaxies,
as described by Rees (1984), and that these objects then grow to quasars
by the accretion of hot gas.

\section{Quasar birth and death} \label{bmodel}

In order to test the outcome of this quasar formation model, we have
incorporated it into a semi-analytical model for galaxy formation, the
details of which are are described in Nulsen \& Fabian (1997) and
Nulsen, Barcons \& Fabian (1998).  In this section we outline
modifications we have made to that model in order to track the growth
and accretion luminosities of nuclear black holes.

As outlined in the previous section, the major factors that determine
the growth rate of a nuclear black hole are the gas temperature and
the Mach number of the cooling flow.  Expressed in terms of the growth
time (\ref{growth}), the time dependence of the mass of a nuclear
black hole is given by
\eqst{mass}
\ln{\Mbh(t_2)\over\Mbh(t_1)} = \int_{t_1}^{t_2} {dt \over
\tg}.
\eqend
The growth time depends on the temperature of the gas, $\Ti$, which is
determined at the time of collapse of a protogalaxy, and the Mach
number, $\machi$, of the isothermal cooling flow.  The latter factor
is the only time dependent part of $\tg$. 

In our galaxy formation model, each collapse produces a dark halo
which is taken to be a perfect isothermal sphere (density $\propto
r^{-2}$) that is truncated at $r_{200}$, the radius within 
which the mean density is 200 times the background density of an
Einstein-de Sitter Universe at the time of collapse.  The gas
temperature produced in the collapse is expressed as
\eqst{beta}
\Ti = {\mu \mh \sigma^2 \over \beta k},
\eqend
where $\sigma$ is the line-of-sight velocity dispersion of the halo
and $\beta$ is a dimensionless parameter, generally lying in the range
0.5 to 1, that allows for excess energy in the gas (mostly excess
binding energy resulting from supernova driven ejection; $\beta=1$
corresponds to zero excess energy; $\beta$ here is determined for each
collapse as in the existing semi-analytical model).  

The outcome of a collapse is determined by considering a notional,
non-radiative collapse.  In this collapse, the gas would form a
hydrostatic atmosphere, with density proportional to $r^{-2\beta}$
(also truncated at $r_{200}$).  The ratio, $\tau$, of the cooling time
to the free-fall time in the notional atmosphere is used to separate
the gas into two parts, one with $\tau < \tau_0$ that is cold (due to
efficient radiative cooling) immediately after the collapse and one
with $\tau > \tau_0$ that forms a hot atmosphere after the collapse.
The model parameter $\tau_0$ is of order unity and determines the
radius, $\rcf$, in the notional atmosphere that separates these two
regions. 

Radiative cooling eventually causes hot gas produced in the collapse
to cool to low temperatures.  As in clusters, the hot gas is little
affected by cooling until the age of the system is comparable to its
cooling time, at which stage it joins a steady cooling flow before
being deposited as cold gas.  Thus the hot atmosphere consists of an
outer region that is largely unperturbed since the collapse, a
transition region, comparable in size to the steady cooling flow, and
a central steady cooling flow.  The total rate of mass deposition and
the extent of the steady cooling flow are determined by the initial
state of the hot atmosphere (\eg Fabian \& Nulsen 1979).  In
particular, this means that the Mach number of the steady cooling flow
is determined by the initial structure of the hot gas.

Formerly, the hot gas was assumed to cool to low temperature at a
time $t = \tcoll + \tcool$, where $\tcoll$ is time of the collapse and
\mst
\tcool(r) = {3\rho(r) k \Ti \over 2 \mu \mh \ne(r)\nh(r) \Lambda(\Ti)}
\propto r^{-2\beta}
\mend
is the cooling time of the hot gas in the notional collapse.  This
made deposition of the cooled gas start discontinuously a short time
after collapse.  In reality, the onset of the cooling flow will be
continuous, commencing immediately after the collapse.  To model this,
here we assume that the gas cools when
\mst
t = \tcoll + \tcool(r) - \tcool(\rcf),
\mend
which gives the radius (in the notional atmosphere) of gas that is
cooling at time $t$ as
\mst
\rcool = \rcf \left[1 + {t - \tcoll \over \tcool(\rcf)}
\right]^{1/2\beta}.
\mend

Since the cooling time of the first hot gas to cool is comparable to
the free-fall time ($\tau_0$ about 1), we should expect the initial
Mach number of the cooling flow, $\machin$, to be about 1.  We treat
this as a parameter of the model.  The time dependence of the Mach
number is then determined from the expectation that the flow velocity
scales with time as $\rcool/\tcool(\rcool) \propto \rcool^{1-2\beta}$,
giving
\eqst{macht}
\machi = \machin \left[ 1 + {t - \tcoll \over \tcool(\rcf)}
\right]^{(1 - 2\beta)/2\beta}.
\eqend
Using this and (\ref{growth}) in (\ref{mass}) enables us to determine the
factor by which the mass of a nuclear black hole grows as a function of the
time.  Differentiating the result with respect to the time gives the
accretion rate and hence the luminosity of the black hole.  Of course,
the black hole stops growing when the hot gas is exhausted ($\rcool >
r_{200}$). 

The low level of emission from nuclear black holes in nearby galaxies
suggests that the radiative efficiency of accretion is much lower now
than it was in quasars (Fabian \& Rees 1995; di Matteo \etal 1999).
Several models, including advection-dominated accretion flows (Narayan
\& Yi 1995), advection-dominated inflow-outflow solutions (Blandford
\& Begelman 1999) and other gas processes (Stone, Pringle \& Begelman
1999), suggest that this is due to a reduced gas supply.  Despite
indications that advection-dominated accretion discs do not account
for the behaviour of some nearby massive black holes (Di Matteo et al
1999), for the sake of definiteness, we base our model on them.  Thus
the radiative efficiency is assumed to be high as long as the accretion
rate exceeds about $1.3 \alphad^2 \mdedd$ (Esin, McClintock \& Narayan
1997).  
When the accretion rate falls below this value, the radiative
efficiency of the nuclear accretion disc is assumed to
plummet, in effect turning off emission from the active nucleus.  This
gives a critical growth time,
\eqst{critgrow}
\tgc = {\Mbh \over 1.3 \alphad^2 \mdedd} 
\simeq 3.5 \times 10^7 \alphad^{-2} {\rm\, y},
\eqend
and when the growth time (\ref{growth}) exceeds $\tgc$, a quasar
turns off, although the nuclear black hole can continue to grow.
$\alphad$ is treated as a parameter of the model.  

Our galaxy formation model uses the block model (Cole \& Kaiser 1988)
to simulate merger trees.  The smallest blocks have a mass of
$1.5\times10^{10} \, \Msun$.  In order to simulate the presence of
seed black holes, a black hole of (arbitrary) unit mass is associated
with each of the smallest blocks.  When a block collapses, the black
holes associated with all merging sub-blocks are assumed to merge into
a single black hole.  This makes the mass of a seed black hole
proportional to the mass of its halo up to the stage that it starts to
grow by Bondi accretion.

A nuclear black hole can only grow by Bondi accretion when it forms in a
collapse that produces some hot gas.  Such systems are identified with
normal galaxies in our model.  The model does not allow mergers between
normal galaxies, so that normal galaxies only grow in collapses where they
accrete dwarf galaxies and gas.  Any collapse involving more than one normal
galaxy is taken to form a group or cluster of galaxies.  Since nuclear black
holes are associated with galaxies rather than a group or cluster, we do not
track the growth of black holes for galaxies in these systems.  In short,
black holes can only grow by Bondi accretion to form quasars in the
``normal'' galaxies of our simulation. McClure et al (1998) find from
optical imaging that most quasars do occur in elliptical, or spheroidal
galaxies, so our model should apply well to such objects. It may not be
directly relevant to present-day low luminosity Seyfert galaxies, which tend
to be in spiral galaxies, but can have undergone a hot-phase era at an
earlier stage.

Note that, since the black hole masses are in arbitrary units, the
quasar luminosities are too.  Adopting a luminosity scale and
radiative efficiency for the quasars will fix the scale of the black
hole masses.  On the bolometric magnitude scale used in the plots in
this paper, $-15$ corresponds to an accretion rate of $10^{-4}$ black
hole mass units per year.  If the radiative efficiency is 0.1 and the
mass unit is taken as $10^4 \,\Msun$, this would correspond to a
bolometric luminosity of about $6\times10^{45} {\rm\, erg\, s^{-1}}$.

\section{Model results}\label{results}

For the purpose of the simulations we have taken an open CDM
cosmology, with $H_0 = 50 {\rm\, km \, s^{-1} \, Mpc^{-1}}$, density
parameter $\Omega = 0.3$, baryon density parameter $\Omega_{\rm b} =
0.075$ and $\sigma_8 = 1$.

Because nuclear black holes grow exponentially in our model, the
results are quite sensitive to the key parameters, $\tau_0$, $\machin$
and $\alphad$.  Using parameter values that favour high growth
produces such massive nuclear black holes by the present day that the
most recent quasars are inevitably the most luminous.  At the other
extreme, parameter values can easily be found that result in
essentially no growth of the seed black holes.  The range of
parameters giving substantial, but not excessive, black hole growth is
relatively narrow (although it covers a substantial part of the
physically reasonably parameter range due to the correlated effects of
the parameters).  Models presented here are chosen to lie in that
range.

\begin{figure}
\centerline{\psfig{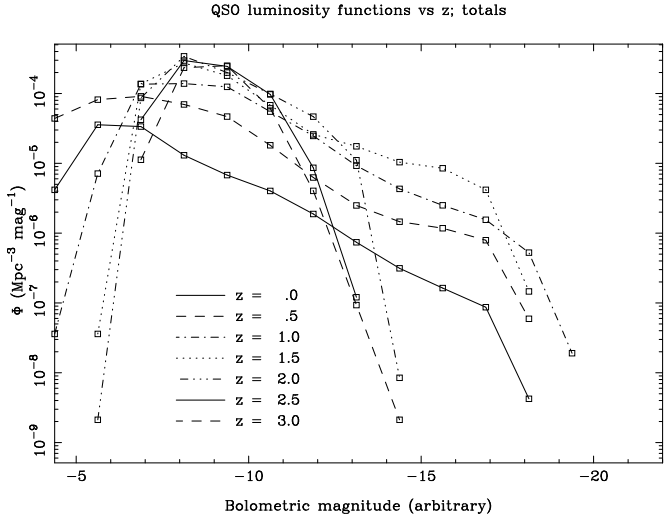}}
\caption{Bolometric luminosity functions for various redshifts.
These are for the model with $\tau_0=1$, $\machi = 0.9$ and $\alphad =
0.1$.  The luminosity used here is the scaled accretion rate in
arbitrary units, as described in section \ref{bmodel}.}
\end{figure}

\begin{figure}
\centerline{\psfig{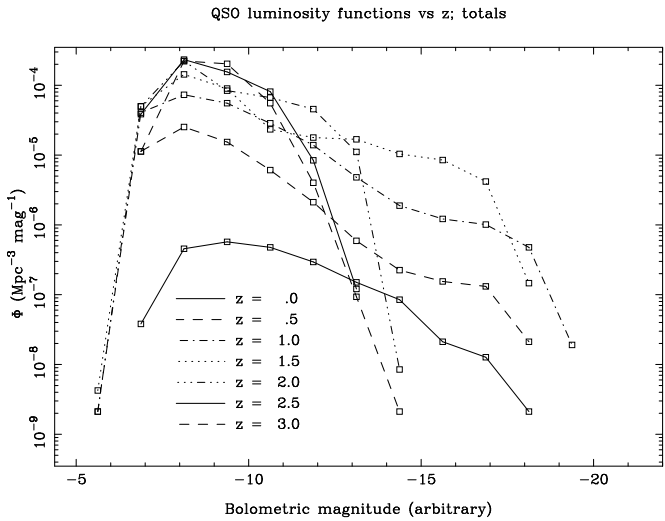}}
\caption{Same as Fig.~1, but for $\alphad=0.15$.}
\end{figure}

Fig.~1 shows distributions of total accretion luminosity for several
redshifts for the case $\tau_0 = 1$, $\machi = 0.9$ and $\alphad =
0.1$, while Fig.~2 shows the same thing for $\alphad = 0.15$.  Since
$\alphad$ only affects the the critical growth time (equation
\ref{critgrow}), black hole masses and accretion rates are identical
in the two models.  Differences between them are entirely due to the
earlier onset of the ADAF phase for the model of Fig.~2.  This effect
is greatest at low redshifts, since a greater proportion of the active
nuclei is then old and cooling flows in the older collapsed systems
have lower Mach numbers (equation \ref{macht}).  A black hole
accreting from a cooling flow with a low Mach number has a longer
growth time (equation \ref{growth}) and so is more likely to be an
ADAF when $\alphad$ is increased.  This accounts for the substantial
reduction in the numbers of luminous active nuclei at low redshifts
between the models of Fig.~1 and Fig.~2.

\begin{figure}
\centerline{\psfig{figure=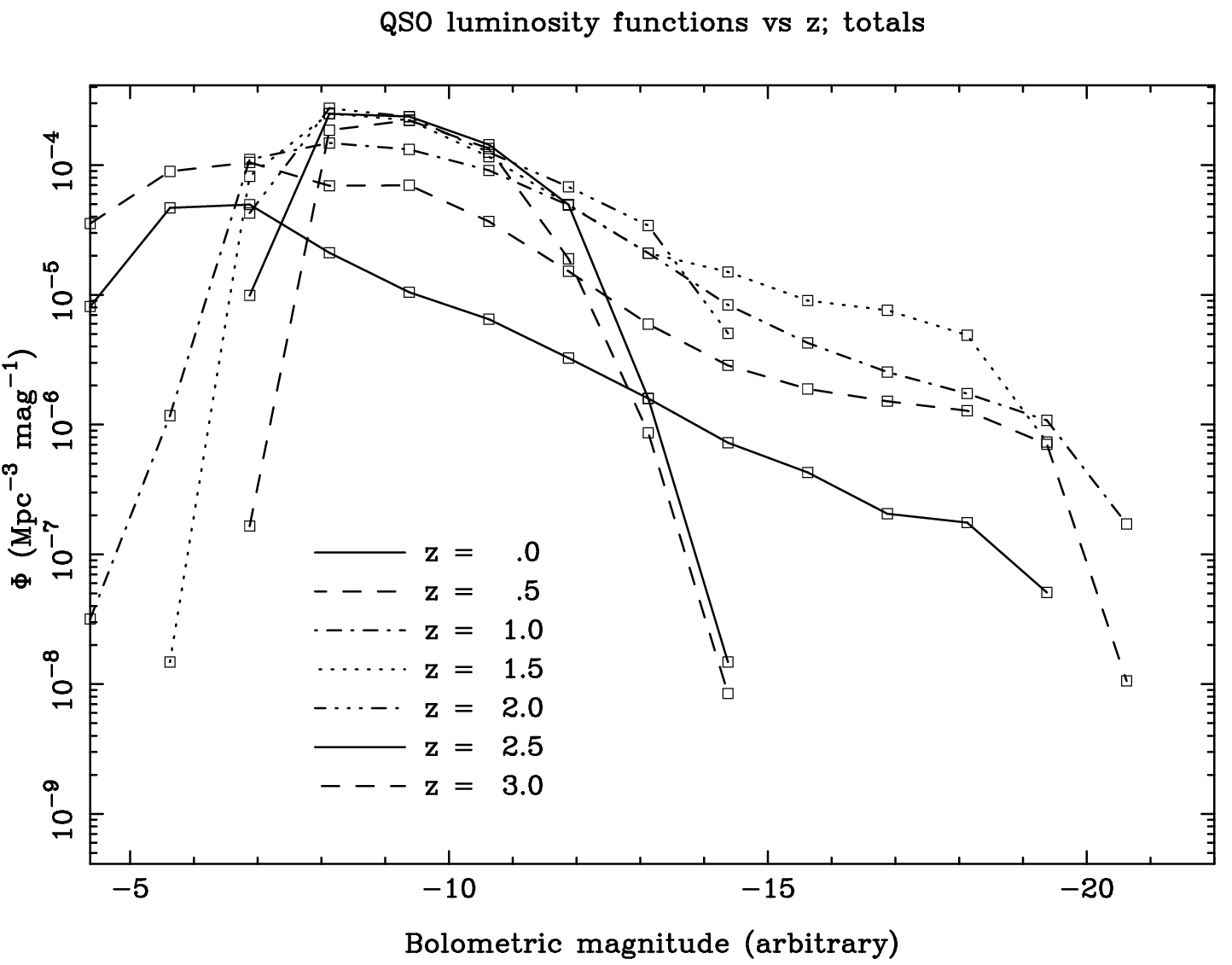,width=0.5\textwidth,angle=270}}
\caption{Same as Fig.~1, but for $\machi = 1$.}
\end{figure}

The model in Fig.~3 is the same as that of Fig.~1, except that the
initial Mach number of the cooling flow is $\machi = 1$.  Increasing
the Mach numbers of the cooling flows increases the nuclear accretion
rate, resulting in greater black hole growth and higher nuclear
luminosities.  This effect can be seen in Fig.~3, where, for the same
seed mass, the most luminous quasars are more numerous at all
redshifts.

\begin{figure}
\centerline{\psfig{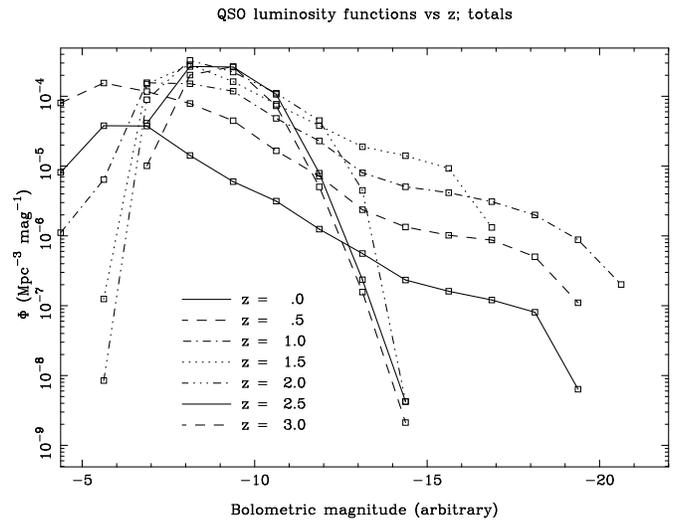}}
\caption{Same as Fig.~3, but for $\tau_0 = 0.9$.}
\end{figure}

Finally, Fig.~4 shows the bolometric luminosity distributions for a
model with $\tau_0 = 0.9$, $\machi = 1$ and $\alphad = 0.1$.  This is
most readily compared to the model of Fig.~3.  The effect of changing
$\tau_0$ is more complicated than that of the other two parameters,
but its main influence on black hole growth is through the cooling
time at the inner edge of the notional hot atmosphere.  $\tau_0$ sets
the ratio of the cooling time to the free-fall time at $\rcf$, the
inner edge of the notional hot atmosphere, so that reducing it reduces
the cooling time there, $\tcool(\rcf)$.  This cooling time sets the
timescale for the evolution of the Mach number of the cooling flow
(equation \ref{macht}).  Reducing $\tcool(\rcf)$ causes the Mach
number of the cooling flow to decrease more quickly, reducing the
overall growth of the nuclear black holes and hence their
luminosities.

The bolometric luminosity is, essentially, just the total accretion
rate of the black holes and not likely to be a good measure of the
visible luminosity of the disc.  Despite its shortcomings, for the
sake of definiteness, we use the thermal disc model (Shakura \& Sunyaev
1973) to estimate the visible luminosity.  This gives the emitted
spectrum
\mst
P_\nu \propto \nu^{1/3} \Mbh^{2/3} \Mdotbh^{2/3},
\mend
for frequency $\nu$.  The resulting ``visible'' luminosity functions
are plotted in Fig.~5, for the quasar formation model of Fig.~1
($\tau_0 = 1$, $\machi = 0.9$, $\alphad = 0.1$).

\begin{figure}
\centerline{\psfig{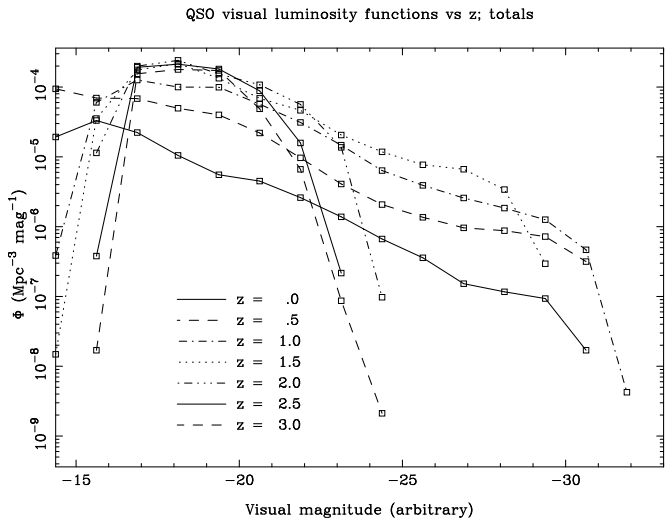}}
\caption{Quasar visible luminosity functions.  This shows visible
luminosity functions corresponding to the bolometric luminosity
functions of Fig.~1, constructed as outlined in the text.  The
magnitude scale is arbitrary.}
\end{figure}

To compare these to the observed quasar luminosity functions (Boyle
\etal 1988), we need to convert our arbitrary magnitude scale to
absolute blue magnitude.  Based on the intensity spectrum of the X-ray
background, Fabian \& Iwasawa (1999) argue that 85 percent of the accretion
power of quasars is absorbed; only ten per cent is seen without some
obscuration at 1~keV.  If so, then the number densities in Fig.~5 (and
preceding Figs) should be reduced by about a factor 10.  In that case,
adding 2 -- 3 to our arbitrary magnitudes to convert to $M_{\rm B}$ gives
rough agreement between our number densities and those of Boyle \etal
(1988).  With this conversion, our luminosity functions are too flat below
about $M_{\rm B} = -27$ and probably too steep above.

The sharp cut offs in the model luminosity functions are largely due
to our assumption that the mass of a seed black hole is proportional
to the mass of the dark halo in which it forms.  A more realistic
model for the seed black holes would give them a distribution of
masses.  In effect, this distribution would be convolved with the
luminosity functions, making them fit the observed luminosity
functions better.

The redshift dependence of our model is also only in rough agreement
with the observed quasar luminosity functions.  The greatest
discrepancy is the absence in our model of luminous active nuclei at
$z=2$ and earlier.  While this is affected to some extent by the
collapse model (\ie cosmology), in large part it is due to the time
required to grow massive black holes.  From equation (\ref{growth}),
the growth rate is maximized by minimizing $\Lambda(\Ti)/ \Ti$, which
generally means in the hottest collapses.  The virial temperature of a
halo of mass $M$ collapsing at time $\tcoll$ scales as
$(M/\tcoll)^{2/3}$, so that, for a given mass, the earliest collapses
give the most growth.  However, few massive galaxies collapse
early, and, since it requires several growth times (equation
\ref{ngrow}) to produce a massive black hole, very few of these form
early in our model.

\begin{figure}
\centerline{\psfig{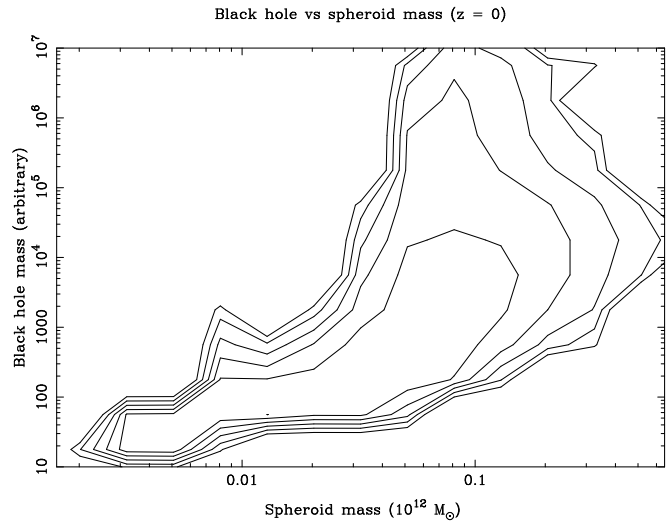}}
\caption{Black hole vs spheroid mass.  This is a contour plot of the
distribution of the masses of the nuclear black holes (arbitrary
units) vs spheroid mass for the model of Fig.~1.  Contours are in
equal logarithmic intervals.}
\end{figure}

Fig.~6 is a contour diagram of the distribution of the masses of the
nuclear black holes vs spheroid mass at $z=0$ in the model of Fig.~1.
Magorrian \etal (1998) and Richstone \etal (1998) find that black hole
mass is proportional to blue luminosity of the spheroid (bulge).
However, there is a substantial spread in this relationship and the
data would be consistent with a substantially steeper relationship for
the more massive bulges.  This is roughly consistent with the ridge
and the lower spur at high spheroid mass of the distribution shown
here.  However, there is no evidence in the data for the extension to
high black hole mass for spheroids of about $10^{11} \,\Msun$ in the
model. 

Taking $10^4$ black hole mass units to correspond to a spheroid mass of
$10^{11} \, \Msun$ and using $\Mbh = 0.005 M_{\rm spheroid}$ (Richstone
\etal 1998) gives the conversion factor of $5\times 10^4
\,\Msun$ per black hole mass unit.  In that case, a bolometric
magnitude of $-15$ in the Figs would correspond to a bolometric
luminosity of about $3\times10^{46} \rm\,erg\,s^{-1}$.  

We can also compare the results of Fig.~1 to the quasar X-ray
luminosity function (Miyaji, Hasinger \& Schmidt 1998).  Using the
black hole mass calibration from above and a fixed bolometric
correction of about 50 for the 0.5--2 keV X-ray luminosity (Elvis
\etal 1994; Fabian \& Iwasawa 1999) means that a magnitude of $-15$ in
Fig.~1 corresponds to a 0.5--2 keV X-ray luminosity of about
$6\times10^{44}\rm\, erg\, s^{-1}$.  Assuming that absorption
reduces the X-ray luminosity function by a factor of about 10, as
above, there is rough agreement between the results in Fig.~1 and the
observed X-ray luminosity function of Miyaji \etal (1998).  However,
the fit suffers from essentially the same problems that we found for
the visible luminosity function.  As in that case, the most serious
problem is the lack of high redshift quasars.

\section{Feedback effects}

Ciotti \& Ostriker (1997) argue that feedback from a quasar will
stifle a cooling flow by heating the cooling gas.  By raising the gas
temperature and reducing its density, this could also dramatically
reduce the Bondi accretion rate.  However, using Ferland's (1996)
CLOUDY and the quasar spectra of Laor et al (1997), we find that the
Compton temperature of radio-loud and radio-quiet quasars is approximately
one and two million K, respectively. This is cooler
than the virial temperature of most of the systems that form
quasars. Furthermore, the gas temperature rises inward in the Bondi
accretion flow, so that Compton feedback from the quasar is likely, at
best, to cool the accreting gas.  If Compton cooling is significant,
then it will almost certainly increase the nuclear accretion rate,
possibly causing it to approach the Eddington limit (Fabian \&
Crawford 1990).

If Compton cooling reduces the temperature of gas near to the nucleus,
the reduction in pressure will result in inflow at speeds comparable
to the sound speed in the uncooled gas.  Since the flow within the
Bondi radius is already roughly sonic, the accretion rate will not be
altered dramatically unless Compton cooling is effective beyond the
Bondi radius.  Taking the Compton cooling time as
$t_{\rm C} = {3\pi \me c^2 r^2 / (\sigmat L_{\rm h})}$,
where $\me$ is the electron mass, $\sigmat$ is the Thomson cross
section and $L_{\rm h}$ is the nuclear luminosity, and taking the
flow time from the Bondi radius, $\rb = G\Mbh/\si^2$, as $t_{\rm BF} =
\rb / \si$, we have
\mst
{t_{\rm C} \over t_{\rm BF}} = {3\pi \me c^2 G\Mbh \over \sigmat
L_{\rm h} \si}
= {3 \me c L_{\rm Edd} \over 4 \mh \si L_{\rm h}}
\simeq 0.8 T_6^{-1/2} \left(L_{\rm h} \over L_{\rm Edd}\right)^{-1},
\mend
where the nuclear luminosity has been put in terms of the Eddington
luminosity, $L_{\rm Edd}$.  This shows that Compton cooling will be
significant outside the Bondi radius for most Eddington-limited active
nuclei.  In terms of our model, for the accretion rate (\ref{mdot}),
if the radiative efficiency of the nuclear accretion disc is $\eta =
0.1 \eta_{-1}$ (and $K=2.5$), then 
\mst
{t_{\rm C} \over t_{\rm BF}} \simeq {\me \Lambda \over \eta
\machi^{3/2} \sigmat \si^3} {\ne\nh\over \rho^2}
\simeq 9 \eta_{-1}^{-1} \machi^{-3/2} T_6^{-3/2} \Lambda_{-23}.
\mend

We find that the ionization parameter of the gas at the accretion
radius, $\xi=L/nr_{\rm B}^2\approx {4\pi\eta}m_{\rm p} c^2 s_i\approx
3\times 10^4.$ Under these circumstances, nuclear radiation keeps the
gas highly photoionized, so that the effective cooling function is
close to pure bremsstrahlung.  The Compton cooling time is then less
than the infall time for gas temperatures exceeding about
$3\times10^6$ K when the Mach number $\machi \simeq 1$.  This means
that the accretion rate may well exceed the Bondi rate in the cases
when it would be highest.  No allowance has been made for this in our
model. 

A further effect which may influence some objects is feedback due to
radio jets.  If the central engine produces jets or outflows which
deposit significant energy near the Bondi radius, then the accretion
rate can be much reduced.  It is not clear how such an effect should
be included in our model at this stage.  If, as suggested by McLure
\etal (1998), the radio loud quasars are those with the largest black
holes, then feedback from radio jets might be responsible for limiting
the growth of the black holes in these systems.

Finally, it has also been suggested that a wind might expel the
surrounding gas when quasars becomes sufficiently luminous (Silk \&
Rees 1998; Fabian 1999).  This would lead to a much closer correlation
between bulge and remnant mass.

\section{Discussion}

The implementation of the quasar formation model used here has a
number of shortcomings.  First, we only follow the growth of black
holes in isolated galaxies.  This discounts growth in groups and
clusters.  Because of the high gas temperature, a central galaxy in a
group could potentially accrete very rapidly.  However, the block
model gives no information about the spatial arrangement of collapsing
objects, so we are unable to identify central galaxies in groups and
clusters.  We may therefore be ignoring the most luminous quasars and
the most massive black holes.

The truncated isothermal potentials used in the model lead to gas
density distributions that are more peaked than in more realistic
collapse models (Navarro, Frenk \& White 1997).  This affects the time
development of the cooling flows, changing the evolution of the Mach
number, and so would affect the time dependence of the nuclear
accretion rate (equation \ref{mdot}).  However, the cooling time
of the hot gas is comparable to the free-fall time in normal galaxy
collapses, so we should still expect immediate onset of a cooling flow
with initial Mach number close to 1 in most cases.  The initial
cooling time controls the rate of change of the Mach number while it
is close to 1 (when the growth rate is largest) and this is comparable
to the collapse time.  Thus, we do not expect such a change to have a
dramatic effect on the results of the simulation.  Beyond this,
is not clear how a more realistic collapse model would alter our
results.

In our simple cooling flow model, the nuclear accretion rate is insensitive
to details of the galactic potential.  However, this may change in a more
realistic model, such as one in which a central star cluster is formed.  In
that case, matter deposited by the cooling flow beyond the Bondi radius
could significantly alter the central potential and so affect the nuclear
accretion rate.

The handling of abundances in our galaxy formation model is very
crude, only allowing for the effects of Type II supernovae and
treating the gas as homogeneous.  Increasing the abundance increases
the cooling function, hence the growth time (equation \ref{growth}),
and so would reduce black hole masses and quasar luminosities.  This
may be significant, since the abundances in some quasars appear to be
very high (\eg Hamann \& Ferland 1993; Ferland \etal 1996).  On the
other hand, the cooling function is considerably less sensitive to
abundance for temperatures exceeding about $3\times10^6$ K and the gas
temperature exceeds this value in most of the systems that would be
quasars, so we should not expect this to have a major effect on the
outcome of the model.

As discussed in section \ref{results}, our assumption that the mass of
a seed black hole is proportional to the mass of halo in which it
resides is too simplistic.  Given that Seyfert nuclei can occur in
disc galaxies, it seems likely that some active nuclei are not fuelled
by hot gas (or, at least, not by gas from a hot halo resulting from
the collapse of the protogalaxy).  A wide variety of other mechanisms
for fuelling active nuclei have been proposed, including starburst
activity, interactions between galaxies and the effects of a bar.
There is also some cold gas within the region that is effectively
drained through a cold accretion disc.  Some or all of these gas
sources may fuel seed black holes, in which case, they could have a
wide range of masses, depending on details of the history of each
galaxy.  Such effects would be compounded with those due to the
processes described in Rees (1984), that are also likely to lead to a
range of seed masses.

If a large proportion of active nuclei are heavily absorbed, then
mergers between galaxies may affect their luminosity by disturbing the
absorbing material, which could alter the luminosity in either
direction. In other words, it is possible that much of the evolution
of observed optical quasars is due to changes in the obscuring
material rather than changes in accretion rate.

The worst shortcoming of our model is its failure to
produce quasars at high redshifts.  For the cosmological parameters
used, a $10^{12} \,\Msun$ halo would have collapsed from a $3\sigma$
peak at about $10^9$ y ($z \simeq 7$).  The gas temperature in a halo
of mass $10^{12} M_{12} \, \Msun$, collapsing at $10^9 t_9$ y, is 
\mst
T \simeq 3.3\times10^6 M_{12}^{2/3} t_9^{-2/3} \beta^{-1} {\rm\, K},
\mend
for $\beta$ as defined in equation (\ref{beta}), so that the growth
time for a nuclear black hole in such a system would have been 
\mst
\tg \simeq 1.5\times10^8 \beta \machi^{-3/2} M_{12}^{-2/3} t_9^{2/3}
\Lambda_{-23} {\rm\, y}.
\mend
The cooling function depends on abundances and ionization (B\"ohringer
\& Hensler 1989), but $\Lambda_{-23}$ lies in the the range 0.5 -- 2
for the relevant temperatures.  Thus, an early collapsing protogalaxy
would have had roughly 10 $e$-folding times to form a massive nuclear
black hole before $z\simeq 3$.  This shows that the lack of high
redshift quasars is not a fundamental shortcoming of our model,
although details of the model would clearly need to be modified in
order to account for them.

The failure of the model to account for the observed relationship
between black hole and spheroid mass (Richstone \etal 1998) is due
primarily to the exponential black hole growth, which tends to make
the larger black holes grow very large.  As discussed in the previous
section, different feedback mechanisms may enhance or limit the growth
rate.  If feedback from a radio source limits the growth of the most
massive black holes, then this, rather than the fuel supply, might be
the cause of the relationship between spheroid and black hole mass.
Note that we have ignored many effects in the nuclear accretion disc
that could break the simple connection we have assumed between the
Bondi accretion rate and the nuclear accretion rate.

Finally, we have attempted to account for the masses of the remnant
nuclear black holes and the evolution of the quasars with a single
mechanism.  If the processes that form the seed black holes account
for a substantial part of their mass, or, if other gas sources also
play a significant role in the fuelling of active nuclei, then this
will not be possible.  In that case, accretion of hot gas may simply
be one of several fuel sources for quasars.  Nevertheless, the results
in section \ref{feed} show that hot gas is potentially a significant
fuel source for AGNs.

Conditions are most favourable for quasar formation in our model when
the hot gas supply is greatest, \ie soon after the collapse of a large
protogalaxy.  In hierarchical collapse models, a collapse will
generally include the infall of gas and other galaxies, so that the
presence of a close companion of comparable luminosity may be
interpreted as an indication of recent collapse.  Thus, the results
that have been interpreted as showing the gravitational interactions
can drive quasars (\eg Bahcall \etal 1997) may also be interpreted as
indicating that quasars form in systems that have collapsed recently,
as expected in the present model.  For this purpose, the main
difference between the predictions of the models is that no close
companion is required in the case that quasars are fuelled by hot gas.

If quasars are fuelled by hot gas, then there should be substantial
halos of hot gas around them.  Since the gas temperature typically
exceeds $3\times10^6$ K, these should be detectable by their X-ray
emission.  The limited extent of the hot gas (comparable to the size
of the dark halo) makes it hard to separate from the powerful nuclear
X-ray emission of a quasar.  Nevertheless, the angular resolution of
Chandra should be sufficient to detect diffuse emission around some
quasars out to redshifts of about 1.

\section{Conclusions}

Bondi accretion of the hot gas produced in the collapse
of protogalaxies onto a seed population of nuclear black holes is sufficient
to form and fuel quasars.  A simple simulation shows that this model can
account for the optical and X-ray luminosity functions of quasars for $z
\approxlt 1.5$, provided that about 90 percent of quasars are obscured.
The simulation produces insufficient quasars at high redshifts and predicts
a wider range of black hole masses in massive spheroids than has been found.

Hot gas formed in the collapse of large protogalaxies is likely to be
the minimum source available for fuelling quasars and so can form the
baseline above which other sources contribute.  Our model directly
confronts and includes problems related to the current fuel supply of
massive black holes in elliptical galaxies. The details of the amplitude and
evolution of hot gas as a fuel supply is sensitive to the presence of
plausible feedback mechanisms, such as heating due to radio jets.

The Bondi accretion rate from the hot gas formed in the collapse of a
protogalaxy can exceed the Eddington accretion rate in systems with
virial temperatures exceeding about $3\times10^6$ K, and can be
enhanced by feedback due to Compton cooling in such systems.
Thus, hot gas is an excellent fuel source for quasars.

\section*{ACKNOWLEDGEMENTS}  PEJN gratefully acknowledges the
hospitality of the Institute of Astronomy, Cambridge, during part of
this work.  ACF thanks the Royal Society for support.

\appendix

\section[]{Bondi accretion from a cooling flow}

We consider Bondi accretion by a massive black hole at the centre of a
steady inhomogeneous cooling flow.  We will show that, for the cooling
flow model used here, the accretion rate onto the black hole depends
on conditions at the edge of the steady cooling flow, but is
insensitive to details of the gravitational potential between there
and the central black hole.  In particular, when the ratio of specific
heats, $\gamma=5/3$, the accretion rate is completely independent of
the intervening potential.

The cooling time of the gas is
\mst
\tcool = {p \over (\gamma-1) \ne\nh \Lambda(T)},
\mend
where $p$ is the pressure, $\ne$ the electron density, $\nh$ the
hydrogen (proton) number density, $T$ the temperature and $\Lambda(T)$
the cooling function.  The flow time of the gas is
\mst
\tflow = {r \over v},
\mend
where $r$ is the radius and $v$ the flow velocity (positive
inward).  For the Bondi solution (\eg Shu 1991) at small $r$,
$v \sim r^{-1/2}$, the density varies as $r^{-3/2}$ and the
temperature as $r^{-3(\gamma-1)/2}$, so that the ratio of the cooling
time to the flow time varies as
\mst
{\tcool\over\tflow} \sim {T\over\Lambda(T)},
\mend
Since the temperature increases as $r$ decreases and, for the relevant
temperatures and abundances, $T/\Lambda$ is almost always an
increasing function of $T$, this ratio increases with decreasing $r$.
In a steady cooling flow $\tcool \simeq \tflow$ (Fabian 1994), but as
the flow comes under the influence of a central black hole
$\tcool/\tflow$ will increase, eventually making cooling
negligible.  Thus, at sufficiently small $r$ cooling can be ignored
and the flow asymptotes to the Bondi solution. 

We begin by outlining the Bondi solution.  For a steady, spherical
flow, the mass flow rate (inward) is
\mst
\dot M = 4\pi \rho v r^2 = {\rm constant},
\mend
where $\rho$ is the gas density.  The flow is adiabatic, so that
\mst
{T \over \rho_{\phantom{0}}^{\gamma-1}} = {T_0 \over \rho_0^{\gamma-1}},
\mend
where $T_0$ and $\rho_0$ are evaluated a long way from the black hole.
Using these, we can express the flow speed as 
\eqst{vbondi}
v = {\dot M \over 4\pi \rho_0 r^2 } \left(T_0\over
T\right)^{1/(\gamma-1)}.
\eqend
Bernoulli's theorem gives
\eqst{benerg}
H + \half v^2 - {GM\over r} = H_0,
\eqend
where $r$ is the radius, $H = \gamma p / [(\gamma-1) \rho]$ is the
specific enthalpy of the gas, $H_0$ is the specific enthalpy at
temperature $T_0$ and $M$ is the mass of the black hole.
Differentiating (\ref{benerg}) with respect to $r$ and using
(\ref{vbondi}) to find $dv/dr$ gives
\mst
\left(H - {v^2\over\gamma-1}\right) {1\over T}{dT \over dr} = {2v^2
\over r} - {GM\over r^2},
\mend
so that at the sonic point, $r_{\rm s}$, where $v_{\rm s}^2 =
(\gamma-1) H_{\rm s}$, we must have
\mst
v_{\rm s}^2 = {GM\over 2 r_{\rm s}}.
\mend
Using these results in (\ref{benerg}) gives
\mst
{GM\over r_{\rm s}} = {4 (\gamma-1) \over 5 - 3 \gamma} H_0,
\mend
enabling us to evaluate all quantities at the sonic point in terms of
$T_0$, $\rho_0$ and $M$.  Evaluating the mass flow rate at the sonic
point then gives the Bondi accretion rate as
\eqst{bondac}
\Mdotbh = \pi \rho_0 {(GM)^2 \over s_0^3 }
\left(5-3\gamma\over 2\right)^{3\gamma-5 \over 2(\gamma-1)},
\eqend
where $s_0 = \sqrt{\gamma p_0 /\rho_0}$ is the speed of sound in gas a
long way from the black hole.  The last factor in (\ref{bondac}),
\mst
q(\gamma) = \left(5-3\gamma\over 2\right)^{3\gamma-5 \over 2(\gamma-1)},
\mend
is finite for both $\gamma\to1$, when $q\to e^{1.5}$, and $\gamma\to
5/3$, when $q\to 1$. 

Now consider a cooling flow.  Thermal instability causes an
inhomogeneous cooling flow to deposit gas throughout the flow at a
rate that is conveniently expressed as (Nulsen 1986, 1988; values
there assume $\gamma=5/3$) 
\mst
\xi {(\gamma-1) \rho R \over \gamma p},
\mend
where $R$ is the power radiated per unit volume by the gas, $\rho$ is
now the mean density of the gas and $\xi$ is dimensionless.  The value
of $\xi$ depends on details of the density distribution and physical
behaviour of the inhomogeneous gas, and is typically of order unity.
We treat it as a constant parameter in what follows, since this leads
to the simplest physically useful flow models (this approximation is
exact for the isothermal cooling flow models; Nulsen 1998).  The mass
conservation equation for a steady flow is then
\eqst{mcons}
{1\over r^2} {d\over dr} \rho v r^2 = \xi {(\gamma-1) \rho R \over
\gamma p}.
\eqend
The corresponding energy equation is (Nulsen 1988)
\eqst{econs}
{1\over \gamma-1} p v {d\over dr} \ln \Sigma = (1 - \xi) R,
\eqend
where $\Sigma = T/\rho^{\gamma-1}$, with $T$ being the temperature
corresponding to the mean density ($\rho$) at pressure $p$.  $\Sigma$
determines the effective entropy of the inhomogeneous gas mixture.
Eliminating $R$ between these two equations and integrating gives 
\eqst{sigmdot}
{\Sigma\over\vphantom{\dot M} \Sigi} = \left(\dot M \over \Mdoti
\right)^{\gamma(1-\xi)/\xi}, 
\eqend
where $\Mdoti$ and $\Sigi$ are evaluated at $\ri$, a fixed point in
the cooling flow, which we will take to be at the outer edge of the
steady flow, well outside the region that is perturbed by the black
hole.  Thus, the cooling flow model enforces a fixed relationship
between the entropy and $\dot M$ throughout the steady flow.

The derivation of equation (\ref{sigmdot}) makes no reference to the
potential (or the momentum equation) and it applies at small $r$, in
the Bondi flow where cooling is negligible.  The quantities $\rho_0$
and $T_0$ (\ie $s_0$) in the Bondi accretion rate (\ref{bondac}) are
no longer well defined, but, since it is constant, the entropy,
$\Sigma_0 = T_0/\rho_0^{\gamma-1}$, is.  Furthermore, (\ref{sigmdot})
requires
\mst
{\Sigma_0\over \vphantom{\dot M} \Sigi} = \left(\Mdotbh \over \Mdoti
\right)^{\gamma(1-\xi)/\xi}, 
\mend
giving
\mst
\rho_0 = \left(T_0 \over \vphantom{\dot M} \Sigi \right)^{1 \over
\gamma-1} \left(\Mdotbh \over \Mdoti \right)^{- { \gamma(1-\xi)
\over \xi(\gamma-1)}}. 
\mend
We use this to eliminate $\rho_0$ in (\ref{bondac}), then solve the
resulting equation for $\Mdotbh$ and put $\Sigi = \Ti /
\rhoi^{\gamma-1}$, where $\Ti$ and $\rhoi$ are the temperature and
density at $\ri$.  After some algebra this gives 
\eqst{invmdot}
\Mdotbh = \left[\pi \rhoi  {(GM)^2\over \si^3} q(\gamma)
\right]^\kappa \Mdoti^{1-\kappa} 
\left(T_0\over \Ti \right)^{\kappa(5 - 3\gamma) \over 2 (\gamma-1)},
\eqend
where $\kappa = \xi (\gamma - 1)/ (\gamma - \xi)$.

This argument makes no reference to the gravitational potential in
which the cooling flow takes place, but $\Mdotbh$ depends on details
of the potential in two ways.  First, gas properties at the edge of
the cooling flow are affected to some extent by the potential
($\Mdoti$ is governed largely by initial conditions in a collapse).
Generally, we can assume that any disturbance to the potential is at
$r \ll \ri$ and has little effect on the cooling flow at $\ri$.  The
second means of influence is through $T_0$, which is still not well
defined.  Despite this, we can expect $T_0$ to be comparable to the
gas temperature at the point where the black hole starts to have an
appreciable affect on the cooling flow.  Since the influence of the
black hole is felt outside the sonic point, this temperature will be
comparable to the ``virial'' temperature at that place and, unless the
galaxy potential is strongly non-isothermal, we can expect it to be
comparable to $\Ti$.  In general, the influence of the potential on
$\Mdotbh$ is weak.

For the case of interest, $\gamma = 5/3$ and $\Mdotbh$ does not depend
on $T_0$, but is determined completely by the mass of the
black hole and the gas properties near the outer edge of the cooling
flow.  This simple result comes about because, for $\gamma=5/3$, the
Bondi accretion rate (\ref{bondac}) depends on the gas properties
through the entropy alone ($\rho_0 s_0^{-3} \propto \Sigma_0^{-3/2}$),
so that it may be regarded as specifying a relationship between
$\Sigma$ and $\dot M$.  The requirement (\ref{sigmdot}) of the cooling
flow specifies a second relationship between $\Sigma$ and $\dot M$
which is only satisfied simultaneously by a unique $\dot M$.

For an inhomogeneous isothermal cooling flow, if $\dot M \propto
r^\eta$, then equations (\ref{mcons}) and (\ref{econs}) require that
\mst
\xi = {2\gamma\eta \over 3(\gamma-1) + \eta (\gamma+1)},
\mend
making
\mst
\kappa = {2\eta\over 3+\eta}.
\mend
We take $\gamma=5/3$ for the remainder of this section.

For an isothermal cooling flow, Nulsen (1998) has shown that
\mst
\rhoi = Q {3-\eta\over2} \left(\rho^2\over \ne \nh \right)
{\vi k \Ti\over \mu\mh \ri \Lambda(\Ti)},
\mend
where $\vi$ the flow speed at $\ri$ and $\rho^2 / (\ne\nh)$ is
constant for the temperatures of interest.  The factor $Q$ is a
constant that depends on details of the cooling function.   For 
power laws, $\Lambda(T) \propto T^a$, with $a$ in the range $[-0.5,
0.5]$, $Q$ ranges from $2.32$ to $2.93$.  Since $d\ln\Lambda/dT$ lies
in about this range for the temperatures of interest, we use the
representative value $Q=2.5$ in all models.  Using the expression for
$\rhoi$ to eliminate the density in $\Mdoti = 4 \pi \rhoi \vi \ri^2$
and using the result to replace $\Mdoti$ in (\ref{invmdot}) gives,
after some further algebra,
\eqst{penul}
\Mdotbh = {(3-\eta) \pi Q k\Ti GM \machi^{3 \over2} \over \mu\mh
\Lambda(\Ti)} {\rho^2\over \ne\nh}  
\biggl({2 \ri \si^2 \machi^{1\over2} \over GM}\biggr)^{3(1-\eta) \over
3 + \eta},  
\eqend
where $\machi = \vi/\si$ is the Mach number at $\ri$.

Observations show that $\eta \simeq 1$ in the well studied cluster
cooling flows (Fabian 1994; Peres \etal 1998) and this is the value
that we use for our models, giving 
\eqst{mainres} \Mdotbh = {2 \pi Q k\Ti GM \machi^{3/2}
\over \mu\mh \Lambda(\Ti)} \left(\rho^2\over \ne\nh\right).  
\eqend 
We note that $\eta$ is not well determined and, since the factor in
parentheses in (\ref{penul}) is usually large, a small change in $\eta$
can have a large effect on $\Mdotbh$.  For the isothermal cooling flow
model, the radial dependence of the Mach number is ${\cal M} \propto
r^{(\eta-1)/2}$, so that $\cal M$ is constant in our models.  If $\eta
< 1$ ($\eta > 1$), then $\cal M$ increases (decreases) inward.  In
general, thermal instability in a cooling flow is greater when the
Mach number is larger (Balbus \& Soker 1989), which will tend to limit
the rise in the Mach number for $\eta < 1$.  This limits the error in
$\Mdotbh$ in that case.  For $\eta > 1$ there is no limiting effect,
so that, if $\eta>1$, (\ref{mainres}) could substantially overestimate
the nuclear accretion rate.  If so, the rate of accretion of hot gas
would be insufficient to account for quasars.

\label{lastpage}
\end{document}